# EMERGENT CONSCIOUSNESS: FROM THE EARLY UNIVERSE TO OUR MIND

P. A. Zizzi

**Dipartimento di Astronomia dell' Università di Padova**
**Vicolo dell' Osservatorio, 5**
**35122 Padova, Italy**
zizzi@giunone.pd.astro.it

In a previous paper (gr-qc/9907063) we described the early inflationary universe in terms of quantum information. In this paper, we analize those results in more detail, and we stress the fact that, during inflation, the universe can be described as a superposed state of quantum registers. The self-reduction of the superposed quantum state is consistent with the Penrose's Objective Reduction (OR) model.
The quantum gravity threshold is reached at the end of inflation, and corresponds to a superposed state of $10^9$ quantum registers. This is also the number of superposed tubulins-qubits in our brain, which undergo the Penrose-Hameroff's Orchestrated Objective Reduction, (Orch OR), leading to a conscious event.
Then, an analogy naturally arises between the very early quantum computing universe, and our mind.

# 1. INTRODUCTION

What is consciousness? Everybody knows about his own consciousness, but it is impossible to communicate our subjective knowledge of it to others. Moreover, a complete scientific definition of consciousness is still missing. However, quite recently, it has been realized that the study of consciousness should not be restricted to the fields of cognitive science, philosophy and biology, but enlarged to physics, more precisely, to quantum physics.

The most popular (and conventional) description of consciousness is based on the classical computing activities in the brain's neural networks, correlated with mental states. In that picture, mind and brain are identified, and are compared to a classical computer. That approach (see for example [1]) is called in various ways: physicalism, reductionism, materialism, functionalism, computionalism.

However, although the brain can actually support classical computation, there is an element of consciousness which is non-computable (in the classical sense), as it was shown by Penrose [2]. Moreover, the seminal paper by Stapp [3] clarified why classical mechanics cannot accommodate consciousness, but quantum mechanics can. Finally, reductionism cannot explain the "hard problem" of consciousness, which deals with our "inner life", as it was illustrated by Chalmers [4].

A quite different line of though about consciousness is the one which comprises pamnpsychism, pan-experientalism, idealism, and funda-mentalism.

Pan-experientalism states that consciousness (or better proto-consciousness) is intrinsically unfold in the universe, and that our mind can grasp those proto-conscious experiences. This line of though goes back to Democritus, Spinoza [5], Leibniz [6], until Withead [7] who re-interpreted the Leibniz's "monads" as "occasions of experience". Shimony [8] compared Withead's occasions of experience to quantum jumps.

More recently, Penrose interpreted the occasions of experience as the quantum state reductions occurring at the Planck scale, where spin networks [9] encode proto-consciousness. This is a pan-experiental approach to consciousness which is consistent with quantum gravity, and is called "Objective Reduction" (OR) [2]. A further development is the Penrose-Hameroff "Orchestrated Objective Reduction" (Orch OR) [10] which deals with the self-collapse of superposed tubulins in the brain. Superposed tubulins are qubits, and perform quantum computation, until they reach the quantum gravity threshold, then they collapse to classical bits, giving rise to a conscious event.

Finally, Chalmers [4] claimed that physical systems which share the same organization will lead to the same kind of conscious experience (Principle of Organizational Invariance). As physical systems which have the same organization (no matter what they are made of ) encompass the same information, it follows, from the above principle, that information is the source of consciousness.

The present paper is on the same line with Chalmers' conclusions.

This is of course valid also for an immaterial system, like the vacuum-dominated early inflationary universe which, as it was shown in [11], is a superposed quantum state of qubits.

At this point, a conjecture arises very naturally: the early universe had a conscious experience at the end of inflation, when the superposed quantum state of $n = 10^9$ quantum gravity registers underwent Objective Reduction. The striking point is that this value of n equals the number of superposed tubulins-qubits in our brain, which undergo Orchestrated Objective Reduction, leading to a conscious event. Then, we make the conjecture that the early universe and our mind share the same organization, encompass the same quantum information, and undergo similar conscious experiences. In other words, consciousness might have a cosmic origin, with roots in the pre-consciousness ingrained directly from the Planck time.

The paper is organized as follows. In Sect.2, we describe the early inflationary universe as a superposed state of quantum gravity registers.

In Sect.3, we show that a quantum gravity register follows some cybernetic principles. From the principles of autopoiesis and self-reproduction, together with the no-cloning theorem, and Chalmers' Principle of Organizational Invariance, we derive the "Principle of Alternating Computational Modes" which begets consciousness.

In Sect.4, we claim that the superposed state of quantum gravity registers undergoes self-reduction as in the Penrose's OR model at the end of inflation, and show that this fact is responsible for the actual entropy of our universe.

In Sect.5, we shortly review the Orch OR model, pointing out that the number of tubulins involved in the Orch OR, equals the number of quantum gravity registers, involved in the OR, and we make the conjecture that the universe might have achieved consciousness at the end of inflation.

In Sect.6, we interpret the Boolean observer as a necessary product of the post-inflationary universe. Moreover, we show that the functor Past can be defined only once a particular quantum gravity register is selected. Sect.7 is devoted to some concluding remarks.

## 2. QUANTUM GRAVITY REGISTERS

A quantum memory register is a system built of qubits. We will consider a quantum register of n qubits.

The state of n qubits is the unit vector in the $2^n$-dimensional complex Hilbert space:
$C^2 \otimes C^2 \otimes ... \otimes C^2$    n times.

As a natural basis, we take the computational basis, consisting of $2^n$ vectors, which correspond to $2^n$ classical strings of length n:

$|0\rangle \otimes |0\rangle \otimes ... \otimes |0\rangle \equiv |00...0\rangle$
$|0\rangle \otimes |0\rangle \otimes ... \otimes |1\rangle \equiv |00...1\rangle$
.
.
.
$|1\rangle \otimes |1\rangle \otimes ... \otimes |1\rangle \equiv |11...1\rangle$

For example, for n=2 the computational basis is:
$|0\rangle \otimes |0\rangle \equiv |00\rangle$

$$|0\rangle \otimes |1\rangle \equiv |01\rangle$$
$$|1\rangle \otimes |0\rangle \equiv |10\rangle$$
$$|1\rangle \otimes |1\rangle \equiv |11\rangle$$

In general, we will denote one basis vector of the state of n qubits as:
$$|i_1\rangle \otimes |i_2\rangle \otimes ... \otimes |i_n\rangle \equiv |i_1 i_2 ... i_n\rangle \equiv |i\rangle$$

where $i_1, i_2, ..., i_n$ is the binary representation of the integer i, a number between 0 and $2^{n-1}$. In this way, the quantum memory register can encode integers.

The general state is a complex unit vector in the Hilbert space, which is a linear superposition of the basis states:

$$\sum_{i=0}^{2^n-1} c_i |i\rangle$$

where $c_i$ are the complex amplitudes of the basis states $|i\rangle$, with the condition:

$$\sum_i |c_i|^2 = 1$$

For example, the most general state for n=1 is: $c_0|0\rangle + c_1|1\rangle$ with $|c_0|^2 + |c_1|^2 = 1$.

The uniform superposition $\frac{1}{\sqrt{2}}(|0\rangle + |1\rangle)$ is the one we will consider in the following, for the n=1 qubit.

To perform computation with qubits, we have to use quantum logic gates. A quantum logic gate on n qubits is a $2^n \times 2^n$ unitary matrix U.

The unitary matrix U is the time evolution operator which allows to compute the function f from n qubits to n qubits:

$$|i_1 i_2 ... i_n\rangle \to U|i_1 i_2 ... i_n\rangle = |f(i_1 i_2 ... i_n)\rangle$$

The hamiltonian H which generates the time evolution according to Schrodinger equation, is the solution of the equation:

$$U = \exp(-\frac{i}{\hbar}\int H dt) \qquad \text{with } UU^+ = I$$

In the following, we will use the same notations of [11], and we will indicate the n=1 qubit as the uniform superposition: $|1\rangle = \frac{1}{\sqrt{2}}(|on\rangle + |off\rangle)$.

In our case, the quantum register grows with time. In fact, at each time step $t_n = (n+1)t^*$ [12] (with n=0,1,2...), where $t^* \cong 5.3 \times 10^{-44}$ sec is the Planck time, a Planckian black hole, (the n=1 qubit state $|1\rangle$ which acts as a creation operator [11]), supplies the quantum register with extra qubits.

At time $t_0 = t^*$ the quantum gravity register will consist of 1 qubit:

$$(|1\rangle|0\rangle)^1 = |1\rangle$$

At time $t_1 = 2t^*$ the quantum gravity register will consist of 4 qubits:

$(|1\rangle|1\rangle)^2 = |2\rangle|2\rangle = |4\rangle$

At time $t_2 = 3t^*$ the quantum gravity register will consist of 9 qubits:

$(|1\rangle|2\rangle)^3 = |3\rangle|3\rangle|3\rangle = |9\rangle$

At time $t_3 = 4t^*$, the quantum gravity register will consist 16 qubits:

$(|1\rangle|3\rangle)^4 = |4\rangle|4\rangle|4\rangle|4\rangle = |16\rangle$

At time $t_4 = 5t^*$, the quantum gravity register will consist of 25 qubits:

$(|1\rangle|4\rangle)^5 = |5\rangle|5\rangle|5\rangle|5\rangle|5\rangle = |25\rangle$

and so on.

The states $|1\rangle$, $|2\rangle$, $|3\rangle \ldots |n\rangle \ldots$ are the uniform superpositions:

$|1\rangle = \frac{1}{\sqrt{2}}(|on\rangle + |off\rangle)$

$|2\rangle = \frac{1}{2}(|on\ on\rangle + |on\ off\rangle + |off\ on\rangle + |off\ off\rangle)$

$|3\rangle = \frac{1}{2\sqrt{2}}(|on\ on\ on\rangle + |on\ ono\ ff\rangle + |on\ off\ on\rangle + |on\ off\ off\rangle + |off\ on\ on\rangle + |off\ on\ off\rangle +$

$+ |off\ off\ on\rangle + |off\ off\ off\rangle)$

and so on.

The general state $|n\rangle$ is:

$|n\rangle = \frac{1}{\sqrt{2}^N} |1\rangle^{\otimes n}$

At time $t_n = (n+1)t^*$ the quantum gravity register will consist of $(n+1)^2$ qubits:

$(|1\rangle|n\rangle)^{n+1} = |n+1\rangle^{n+1} = |(n+1)^2\rangle$

We call $|N\rangle$ the state $|(n+1)^2\rangle$, with $N = (n+1)^2$.

Now, let us consider a de Sitter horizon $|\Psi(t_n)\rangle$ [11] [12] at time $t_n = (n+1)t^*$, with a discrete area $A_n = (n+1)^2 L^{*2}$ [12] (where $L^* \cong 1.6 \times 10^{-33} cm$ is the Planck length) of $N$ pixels.

By the quantum holographic principle [11], we associate N qubits to the $n^{th}$ de Sitter horizon:

$|N\rangle \equiv |\Psi(t_n)\rangle$.

Let us remember that $|1\rangle = Had|0\rangle$ where $Had$ is the Hadamard gate (which is a very important gate for quantum algorithms):

$Had = \frac{1}{\sqrt{2}}\begin{pmatrix} 1 & 1 \\ 1 & -1 \end{pmatrix}$

and $|0\rangle$ is the vacuum state, which can be identified either with the basis state $|on\rangle$ or with the basis state $|off\rangle$.

In fact, let us represent the basis states $|on\rangle$ and $|off\rangle$ as the vectors $\begin{pmatrix}1\\0\end{pmatrix}$ and $\begin{pmatrix}0\\1\end{pmatrix}$ respectively.

The action of *Had* on the vacuum state $|0\rangle \equiv |off\rangle$ is:

$$Had|0\rangle = \frac{1}{\sqrt{2}}\begin{pmatrix}1 & 1\\1 & -1\end{pmatrix}\begin{pmatrix}0\\1\end{pmatrix} = \frac{1}{\sqrt{2}}\left[\begin{pmatrix}1\\0\end{pmatrix} - \begin{pmatrix}0\\1\end{pmatrix}\right] = |1\rangle^A$$, where "A" stands for "antisymmetric".

The action of *Had* on the vacuum state $|0\rangle \equiv |on\rangle$ is:

$$Had|0\rangle = \frac{1}{\sqrt{2}}\begin{pmatrix}1 & 1\\1 & -1\end{pmatrix}\begin{pmatrix}1\\0\end{pmatrix} = \frac{1}{\sqrt{2}}\left[\begin{pmatrix}1\\0\end{pmatrix} + \begin{pmatrix}0\\1\end{pmatrix}\right] = |1\rangle^S$$, where "S" stands for "symmetric".

In the following, we will consider the vacuum state $|0\rangle \equiv |on\rangle$, and the symmetric state $|0\rangle^S \equiv |1\rangle$.

Then, the state $|N\rangle = |(n+1)^2\rangle$ can be expressed as:

$$|N\rangle = (Had|0\rangle \frac{1}{\sqrt{2}^n}|1\rangle^n)^{n+1} = (Had|0\rangle)^{(n+1)^2} = (Had|0\rangle)^N$$

As time is discrete, there will be no continuos time evolution, therefore there will not be a physical Hamiltonian which generates the time evolution according to Schrodinger equation. In [11] we considered discrete unitary evolution operators $E_{nm}$ between two Hilbert spaces $H_n$ and $H_m$ associated with two causally related "events" $|\Psi_n\rangle$ and $|\Psi_m\rangle$. These "events" are de Sitter horizon states at times $t_n$ and $t_m$ respectively, with the causal relation: $|\Psi_n\rangle \leq |\Psi_m\rangle$, for $t_n \leq t_m$.

The discrete evolution operators: $E_{nm} = |1\rangle^{(m-n)(m+n+2)} : H_n \to H_m$.

are the logic quantum gates for the quantum gravity register.

Thus we have: $E_{nm} = |1\rangle^{(m-n)(m+n+2)} \equiv (Had|0\rangle)^{(m-n)(m+n+2)}$, and the discrete time evolution is:

$$E_{0n}|0\rangle = (Had|0\rangle)^{n(n+2)}|0\rangle = |1\rangle^{n(n+2)}|0\rangle = |1\rangle^{n(n+1)}|1\rangle|0\rangle = |1\rangle^{n(n+1)}|1\rangle = |1\rangle^{(n+1)^2} = |N\rangle = |\Psi_{fin}\rangle$$

As the time evolution is discrete, the quantum gravity register resembles more a quantum cellular automata than a quantum computer. Moreover, the quantum gravity register has the peculiarity to grow at each time step (it is self-producing).
If we adopt an atemporal picture, then the early inflationary universe can be interpreted as an ensemble of quantum gravity registers in parallel:

$$|\Psi\rangle = \sum_n \alpha_n |\Psi_n\rangle$$

which reminds us of the many-worlds interpretation [13].
Some related papers on the issue of a quantum computing universe can be found in [14].

## 3. QUANTUM GRAVITY COMPUTATION AND CYBERNETIC

As we pointed out in Sect.2, the quantum gravity registers are not proper quantum computers. Basically, they resemble quantum cellular automata, as time is discrete. Quantum gravity registers do perform quantum computation, but in a rather particular way, that we shall call *quantum gravity computation* (QGC). The peculiarity of a quantum system which performs QGC, is that it shares some features of self-organizing systems. We recall that self-organization is a process of evolution taking place basically inside the system, with minimal or even null effect of the environment. In fact, the dynamical behaviour of a quantum gravity register follows some cybernetic principles:

**i) Autocatalytic growth**
At each computational time step, the presence of a Planckian black hole (which acts as a creation operator), makes the quantum gravity register grow autocatalytically.
As N qubits represent here a de Sitter horizon with an area of $N$ pixels, the autocatalytic growht, in this case, is exponential expansion, i.e., inflation.

**ii) Autopoiesis (or self-production)**
The quantum gravity register produces itself. The components of the quantum gravity register generate recursively the same network of processes (applications of the Hadamard gate to the vacuum state) which produced them.
In this case recursion is defining the program in such a way that it may call itself:
$|N\rangle = (Hadamard|0\rangle)^N$.
This is on the same line of thought of Kauffmann's "Fourth Law" [15]: "…The hypothesis that the universe as a whole might be a self-constructing coevolving community of autonomous agents that maximizes the sustainable growth…".
For Kauffmann, the autonomous agents are knotted structures created of spin networks which act on one another and become collectively autocatalytic. The picture given in this paper and the Kauffmann' s picture, are equivalent, because spin networks pierce the de Sitter horizons' surfaces [12].

**iii) Self-similarity**
This model of the early inflationary universe is based on the holographic principle [16], more in particular, on the quantum holographic principle [11] [17]. But each part of a hologram carries information about the whole hologram. So, there is a physical correspondence between the parts and the whole.

**iv) Self-reproduction**
Can a quantum gravity register, as a unit, produce another unit with a similar organization? This possibility, which could be taken into account because the quantum gravity register is an autopoietic system, (and only autopoietic systems can self-reproduce), is in fact forbidden by the no-cloning theorem [18] (quantum information cannot be copied).

However, there is a way out. When the selected quantum gravity register collapses to classical bits, it is not just an ordinary quantum register which collapses, but an autopoietic one. The outcomes (classical bits) carry along the autopoiesis. The resulting classical automaton is then autopoietic and, in principle, can self-reproduce. Moreover, as it was pointed out in the Introduction, the Chalmers' Principle of Organizational Invariance would assign to the (produced) unit with similar organization, the same amount of information, and the same conscious experience of the original one.

From the cybernetic principles, the Organizational Invariance Principle and from the no-cloning theorem, we get *The principle of alternating computational modes*: "A unit produced by an autopoietic classical computing system built up from the autcomes of a decohered quantum autopoietic system, shares the same organization, the same amount of information, and the same conscious experience of the producing unit. Moreover, in order to share the same conscious experience of the decohered quantum system, the produced unit must alternate quantum and classical computational modes at least once".

The above arguments are summarized in the following scheme:

**Autopoietic quantum register → no-cloning theorem → no self-reproduction → decoherence → autopoietic classical cellular automaton → self-reproduction → produced unit with the same organization → principle of organizational invariance → the produced unit shares the same information content, and the same conscious experience → the produced unit gets both quantum and classical computational modes, the former from the autopoietic quantum register, the latter from the autopoietic classical cellular automata → the modes alternate to each other.**

Then, we are lead to make the conjecture that the final outcome of a quantum gravity register might be a brain. In fact, tubulins in the brain alternate classical and quantum computational modes [10].

A related paper on the issue of a cybernetic approach to consciousness can be found in [19].

## 4. OBJECTIVE REDUCTION AND DECOHERENCE

The superposed state of quantum gravity registers represents the early inflationary universe which is a closed system. Obviously then, the superposed quantum state cannot undergo environmental decoherence.

However, we know that at the end of the inflationary epoch, the universe reheated by getting energy from the vacuum, and started to be radiation-dominated becoming a Friedmann universe.

This phase transition should correspond to decoherence of the superposed quantum state. The only possible reduction model in this case is self-reduction, as in the OR model [2] which invokes quantum gravity.

## 4.1 The discrete energy spectrum and the quantum gravity threshold

The discrete energy spectrum[11] [12] of the de Sitter horizon states at times $t_n = (n+1)t^*$, (with n=0,1,2,...) is: $E_n = \dfrac{E^*}{n+1}$.

where $E^* \cong 1.2 \times 10^{19} \, GeV$ is the Planck energy.

N qubits are associated with the $n^{th}$ de Sitter horizon with area of N pixels, with $N = (n+1)^2$.

The gravitational entropy is: $S_G = \dfrac{1}{4} Area / L^{*2} = \dfrac{1}{4} N \, L^{*2} / L^{*2} = \dfrac{1}{4} N$.

The quantum entropy is: $S_Q = N \ln 2 = (n+1)^2 \ln 2$.

Thus, during inflation, gravitational entropy and quantum entropy are mostly equivalent: $S_G \approx S_Q \equiv S$.

Moreover, as the expression of the quantized cosmological constant [12] is:

$$\Lambda_N \approx \dfrac{1}{N \, L^{*2}},$$

we have: $\Lambda_N \approx \dfrac{1}{S \, L^{*2}}$.

The value of the cosmological constant now is then $\Lambda_N \cong 10^{-56} \, cm^{-2}$ in agreement with inflationary theories.

If decoherence of N qubits occurred now, at $t_{now} = 10^{60} t^*$, (that is, $n = 10^{60}$, $N = 10^{120}$), there would be a maximum gravitational entropy: $S_{G\,MAX} = \dfrac{1}{4} Area / L^{*2} = \dfrac{1}{4} N \approx 10^{120}$.

Of course, we would get the same value, by considering the quantum entropy:
$S_{Q\,MAX} = N \ln 2 \approx 10^{120}$

In fact, the actual entropy is about $S_{now} = 10^{101} \sim 10^{102}$ [2]. This means that decoherence should have taken place when the gravitational entropy was $S_{decoh} \approx 10^{19} \sim 10^{18}$ so that:

$$\dfrac{S_{MAX}}{S_{decoh}} = S_{now} \approx 10^{101} \sim 10^{102}.$$

As $n \sim \sqrt{N}$, decoherence should have occurred for $\bar{n} = 10^9$, which corresponds to the time $t_{\bar{n}} = 10^9 t^* = 10^{-34} \, sec$, which is the decoherence time. (It should be noted that $t_{\bar{n}} = 10^{-34} \, sec$ is, according to inflationary theories, the time when inflation ends).

For $\bar{n} = 10^9$, the corresponding energy is $E_{\bar{n}} = \dfrac{E^*}{10^9} = 10^{11} \, Gev$, which is the quantum gravity threshold. (It should be noted that $E_{\bar{n}} = 10^{11} \, Gev$ is the reheating temperature at the end of inflation).

Why we call the $E_{\bar{n}} = 10^{11} \, Gev$ the quantum gravity threshold?

Let us consider the time interval: $\Delta t = t_n - t_0 = nt_0 \equiv nt^* \approx t_n$.

During the time interval $\Delta t$, the variation in energy is: $\Delta E = E_0 - E_n = \frac{nE_0}{n+1} = nE_n \approx E_0$,

so that: $\Delta E \Delta t \approx n E_n t_n \approx E_0 t_n \approx n E_0 t_0 = n E^* t^*$.

As, for a Planckian black hole, it holds: $E^* t^* \approx \hbar$, we get: $\Delta E \Delta t = n\hbar$ which is the discrete time-energy uncertainty relation centered around a Planckian black hole.

From the equations above it follows: $E_n = \frac{\hbar}{t_n}$

For the particular value $\bar{n} = 10^9$, corresponding to the decoherence time:

$T = t_{\bar{n}} = 10^{-34}$ sec, we have the Penrose quantum gravity threshold: $E = \frac{\hbar}{T} = 10^{11} Gev$.

At this stage the superposed state $|\Psi\rangle = \sum_n \alpha_n |\Psi_n\rangle$ self-reduces to the state $|\Psi_{n=10^9}\rangle$, a quantum gravity register in the superposed state of $10^{18}$ qubits.

### 4.2 Entanglement with the environment

The superposed state of $10^{18}$ qubits will collapse to classical bits by getting entangled with the emergent environment (radiation).

This entanglement process with the environment can be interpreted as the action of a XOR (or controlled NOT) gate, as it was illustrated in [11], which gives the output of the quantum computation in terms of classical bits: the source of classical information in the post-inflationary universe.

### 4.3 Holography and Cellular Automata

Cellular automata (CA) were originally conceived by von Neumann [20], to provide a mathematical framework for the study of complex systems.

A cellular automata is a regular spatial lattice where cells can have any of a finite number of states. The state of a cell at time $t_n$ depends only on its own state and on the states of its nerby neighbors at time $t_{n-1}$ (with $n \in Z$). All the cells are identically programmed. The program is the set of rules defining how the state of a cell changes with respect of its current state, and that of its neighbours.

It holds that the classical picture of holography (given in terms of classical bits) can be described by a classical CA.

**States**: 0 or 1 ("off" or "on").

**Neighbours**: n at each time step $t_n = (n+1)t^*$ (this CA is autopoietic and grows with time).

**Rules**: As there are two possible states for each cell, there are $2^{n+1}$ rules required.

- $t_0 = t^*$     •     2 rules: $\begin{matrix} 1 \to 1 \\ 0 \to 0 \end{matrix}$

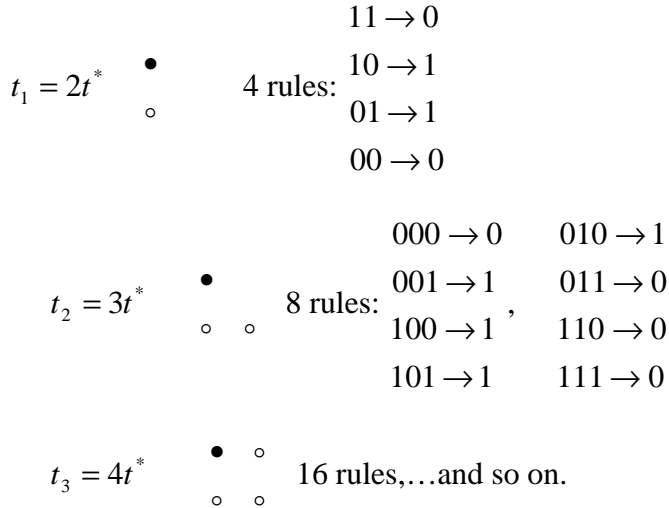

$t_1 = 2t^*$  •  4 rules:  
○  
$$11 \to 0$$
$$10 \to 1$$
$$01 \to 1$$
$$00 \to 0$$

$t_2 = 3t^*$  •  8 rules:  
○ ○  
$$\begin{array}{ll}000 \to 0 & 010 \to 1\\ 001 \to 1 & 011 \to 0\\ 100 \to 1 & 110 \to 0\\ 101 \to 1 & 111 \to 0\end{array}$$

$t_3 = 4t^*$  • ○  
○ ○  16 rules,…and so on.

The rules force patterns to emerge (self-organization).
By taking into account the "classical" holographic principle, we are lead to believe that at the end of inflation, the universe starts to behave as a classical CA which self-organizes and evolves complexity and structure. We call it "Classical Holographic Cellular Automata" (CHCA).
It should be noted that the CHCA is made out of the bits which are the outcomes of the collapse of the qubits of the quantum gravity register which is an autopoietic quantum system. Then, the CHCA is an autopoietic classical system.
There are two important consequences.
i) The CHCA, beeing autopoietic, undergoes autocatalytic growth, and the classical universe is still expanding. However, as classical computation is slower than quantum computation, the expansion is not anymore exponential (post-inflationary universe).
ii) The CHCA, beeing a classical autopoietic system, can self-reproduce.
According to the "Principle of alternating computational modes" discussed in Sect.3, the produced units will be able to perform both quantum and classical computation.
We conclude by saying that in our model, the post-inflationary, classically holographic universe, follows the laws of classical complex adaptive systems (systems at the edge of chaos).

## 5. CONSCIOUSNESS AND TUBULINS/QUBITS

So far, consciousness was studied in the context of neuroscience, and was described as an emergent feature of classical computing in the brain's neural networks.
But neuroscience fails to explain some features of consciousness as, for example, subjective experience (Chalmers' "hard problem" [4]).
A new, different approach to the study of consciousness is due to Penrose and Hameroff [10] and it is based on quantum effects occurring in tubulins.
In a brain' s neuron there is the cytoskeleton, which is made of protein networks. The most important components of the cytoskeleton are microtubules. Microtubules are hollow cylindrical polymers of proteins called tubulins.Tubulins are dipoles and they can be in (at least) two different states (or conformations).

Tubulins have been studied in classical computing. In fact simulations suggest that tubulins behave as a classical CA.
But tubulins can also be in a superposition of the two (or more) conformation states. In this case they are qubits, and they behave as a biological quantum cellular automata. Indeed, tubulins can perform both classical and quantum computing.
In a classical computing mode, patterns of tubulins move, evolve, interact and lead to new patterns.
Quantum coherence emerges from resonance in classical patterns.
When the quantum gravity threshold is reached, self-collapse occours and the eigestate evolves as a classical CA.
In the orchestrated objective reduction (Orch OR) model by Penrose and Hameroff [10], the number of tubulins/cell involved in the threshold is $n = 10^9$, with a coherence time T=500 mse.
As tubulins are qubits, we can indulge in speculating about the brain-universe, with $n = 10^9$ quantum gravity registers, and a coherence time $T = 10^{-34}$ sec, which might have a conscious experience.
Then, if the inflationary universe, which performed quantum computation, was able to achieve consciousness, so will do any quantum computer?
For the moment, the only possible answer is no, for three reasons:
1. Because quantum computers are very difficult to build in practice, as the technology is not yet so advanced to maintain coherence for a sufficiently long time.
2. Because quantum computers (at least the first generations) will not have enough mass.
3. For a quantum system to be able to get a conscious experience, it is a necessary but non sufficient condition that it performs quantum computation. The extra requirement is that the quantum computing system should be quantum-autopoietic.
However, we cannot really foresee anything: In the long run quantum computers might have conscious experiences.

## 6. CONSCIOUSNESS ARISES IN THE "BITS ERA"

There have been some attempts in appending consciousness to the universe as a whole [21].
In our model, during the "qubits era" there are no events in the usual sense, (occasions of experience, in the philosophical language of Withead [7]) although, there can be events in a non-Boolean sense (some work is in progress on that [22]).
So, if we, Boolean minded beeings, conceive consciousness in terms of occasions of experience (events in the Boolean sense), we can argue that in the qubits era there was no consciousness at all in the universe (peraphs, there was just sub-consciousness). Consciousness appeared in the classical "bits era": it was the projection in the past of future observers who had to be programmed by the self-organizing CA, in order to observe the emergent events.

## 6.1 The Boolean observer

To observe the events in the post-inflationary universe, the observers should be Boolean.
This means that the qubits-tubulins of the observers' brain should collapse to classical bits at a rather high frequency.
Of course, beeing the observers Boolean, they will not be able to grasp the unfolding quantum computing structure of spacetime at the fundamental level (the Planck scale) by the use of tools like causal sets [23] and the functor Past [24], which become useless at that scale, where locality and causality are missing (some work is in progress on this issue [25]).
What the Boolean observers can do, is just to recognize the large scale structure of the universe, and, by the use of the functor Past, go backward in time.
But, anyway, the "travel" will stop at the end of inflation: the big bang will never be reached because then the "multiverse" started, for which the functor Past loses any meaning.
The problem is that a Boolean observer is endowed with the concept of time, which is a mere artifact of his own perception, and moreover, he tends to extend this concept to regions of reality where it is meaningless.

## 6.2 The functor Past and the quantum registers

The functor Past [24] is the functor from a causal set C [23] to the category of sets, Set:
$Past : C \rightarrow Set$
Past has components for each event p of C, which are the sets: $\{r \in C : r \leq p\}$

### i) The causal set interpretation

The de Sitter horizon states $|\Psi_0\rangle, |\Psi_1\rangle, |\Psi_2\rangle, \ldots |\Psi_n\rangle$ (which are N-qubits, with $N = (n+1)^2$) can be considered as the events of a causal set.
The causal structure is enbodied in the quantum entropy: $S_N = N \ln 2$, with:

$$\Delta S = S_M - S_N = (M - N) \ln 2 = (m-n)(m+n+2) \ln 2 = \frac{(t_m - t_n)}{t^*}(m+n+2) \ln 2$$

So that: $\Delta S \geq 0$ for $t_n \leq t_m$
The causal relation holds:
$|\Psi_n\rangle \leq |\Psi_m\rangle$ for $t_n \leq t_m$.
and the properties of reflexivity, antisymmetry and transitivity are satisfied. .
In this context, the Functor Past can be defined, but there is no quantum computation. In fact, as in this case the logic gates are connecting the quantum gravity registers one to another, the quantum gravity registers cannot process information individually.
This picture is strictly related to the idea of an internal observer [24], which is not an adequate assumption as far as the very early universe is considered.

### ii) The Fock space interpretation

Let us consider the linear superposition of de Sitter horizon states $|\Psi_n\rangle$:

$$|\Psi\rangle = \sum_n \alpha_n |\Psi_n\rangle$$

where $|\Psi\rangle$ is the wavefunction of the Fock space $\overset{\infty}{\underset{n=1}{\oplus}} H_n$.

In this atemporal picture, the early inflationary universe is interpreted as an ensemble of quantum registers in parallel, and quantum computation is runned overall.
But the functor **Past** cannot be defined. In fact, this is a many-worlds interpretation [13] and the causal past is not unique [26].

### 6.3 Objective reduction, decoherence and the emergence of Past

The superposed state $|\Psi\rangle$ self-reduces to the state $|\Psi_{n=10^9}\rangle$ which is the only "event" which actually takes place in the causal set of possible events $|\Psi_n\rangle$ and implies the possible presence of a future observer in a world W with entropy $S_W = E^2/10^{18}$, where E indicates the epoch in Planck time units. Our epoch is $E_{now} = 10^{60}$.
(The letter W identifies the only world we know. If the superposed state self-reduced for a different value of n, let us say $n' > n$ the actualized world would be a world W' with smaller entropy, conversely for $n' < n$).
At this point, the functor **Past** can be defined. Thus, although the quantum past is not unique, in the world W the past can be re-constructed univocally.

### 6.4 The analogy
Inflation (the "qubits era") is for the universe what pre-consciousness (superposed tubulins) is for our mind.
The end of inflation (beginning of the "bits era") is for the universe what consciousness (Orch OR of superposed states of tubulins) is for our mind.
The analogy goes like that:
**For tubulins in the brain:**
CLASSICAL CA → EMERGENCE OF QUANTUM COHERENCE (PRE-CONSCIOUSNESS) → QUANTUM CA → SELF-COLLAPSE BY ORCHESTRATED OBJECTIVE REDUCTION → CONSCIOUS EXPERIENCE → CLASSICAL CA.
**For qubits in the early universe:**
CLASSICAL BIT (THE VACUUM) → HADAMARD QUANTUM LOGIC GATE → QUBIT → BEGINNING OF INFLATION (THE UNIVERSE IS A SUPERPOSED STATE OF QUANTUM REGISTERS) → SELF-REDUCTION BY OBJECTIVE REDUCTION (END OF INFLATION) → CONSCIOUS EXPERIENCE → COLLAPSE OF QUBITS TO BITS (THE XOR GATE) → CLASSICAL CA.
Of course, the analogy between our mind and the universe is very speculative at this stage, but the emergent picture is quite exciting: it seems that our brain owes its stucture and organization to the very early universe.
This is in agreement with the Penrose-Hameroff's belief that consciousness is a fundamental property of reality, and has its roots in the spacetime structure at the Planck scale.

Then, although we can be just classical as observers, we can be also quantum as thinkers (for example, we can conceive quantum computation). This fact must be the result of some kind of *imprinting* we received from the quantum computing early universe.
If we had not both quantum and classical computational modes available in our brain, in other words, if we were always conscious and Boolean, we would not be able to think quantum.

## 7. CONCLUSIONS

In this paper, we described the early inflationary universe as an ensemble of quantum gravity registers in parallel.
At the end of inflation, the superposed state self-reduces by reaching the quantum gravity threshold as in the Penrose's Objective Reduction model. This self-reduction can be interpreted as a primordial conscious experience. Actually, the number of quantum gravity registers involved in the OR equals the number of superposed tubulins in our brain, which are involved in the Orch OR, leading to a conscious experience. Further, the qubits of the selected quantum gravity register get entangled with the emergent environment and collapse to classical bits. This environmental collapse is the source of classical information and Boolean logic in the actual universe.
Thus, we make the conjecture that the post-inflationary universe starts to organize itself, very likely as a classical Cellular Automata, and necessarily produces self-similar computing systems (our minds). In this way, the actual universe and its products use the same (Boolean) logic so that the past can be recorded, and information can be stored.
It should be noted that, in this model, the quantum gravity registers in parallel are parallel universes. This interpretation is very much on the same line with Deutsch' idea relating quantum computers to parallel universes (the "multiverse") [27].
However, at the end of inflation, only one universe is selected, the one which is endowed with that particular amount of entropy which makes it our world.

## ACKNOWLEDGMENTS

I am grateful to R. E. Zimmermann for helpful discussions.
I thank the Department of Astronomy, University of Padova, Italy, for hospitality.